\documentstyle[aps,preprint,axodraw]{revtex}
\def\btabl{\begin{table}}   \def\etabl{\end{table}}
\def\bea{\begin{eqnarray}}   \def\eea{\end{eqnarray}}
\def\bnn{\begin{eqnarray*}}   \def\enn{\end{eqnarray*}}
\def\beq{\begin{equation}}   \def\eeq{\end{equation}}  
\def\btabu{\begin{tabular}}   \def\etabu{\end{tabular}}
\def\bec{\begin{displaymath}} \def\eec{\end{displaymath}}
\def\nn{\nonumber}
\def\eqref#1{(\ref{#1})}
\renewcommand{\baselinestretch}{1.2}
\begin{document}
\newcommand{\bfig}{\begin{center}\begin{picture}}
\newcommand{\efig}[1]{\end{picture}\\{\small #1}\end{center}}
\newcommand{\flin}[2]{\ArrowLine(#1)(#2)}
\newcommand{\wlin}[2]{\DashLine(#1)(#2){3}}
\newcommand{\zlin}[2]{\DashLine(#1)(#2){5}}
\newcommand{\glin}[3]{\Photon(#1)(#2){2}{#3}}
\newcommand{\lin}[2]{\Line(#1)(#2)}
\newcommand{\sof}{\SetOffset}
\draft
\preprint{\vbox{\baselineskip=13pt
\rightline{CERN-TH/98-241}
\rightline{LPTHE Orsay-98/43}
\rightline{hep-ph/9808nnn}}}
\title{Direct computation of inelastic photon--neutrino processes in the Standard Model
}
\author{A. Abada, 
J. Matias  and R. Pittau \footnote{
e-mail: abada@mail.cern.ch, matias@mail.cern.ch, pittau@mail.cern.ch.}} 
\vskip -0.5cm
\vspace{-0.5cm}
\address{{\small{\em 
Theory Division, CERN, CH-1211 Geneva 23, Switzerland}}}

\vskip 4cm 

\maketitle \begin{abstract}  
In this paper, we compute the Standard Model polarized amplitudes and 
cross sections of the processes $\gamma\nu\to\gamma\gamma \nu$, 
$\gamma\gamma \to \gamma\nu\bar\nu$ and
$\nu\bar\nu   \to \gamma\gamma\gamma$, for centre-of-mass energies $\omega$ 
within the range of validity of the Fermi theory.

\noindent By using a large electron-mass expansion of the exact result, 
we also derive the first correction term  
to the effective, low-energy ($\omega < m_e$) formula for $\gamma\nu\to\gamma\gamma \nu$.
Finally, we discuss possible astrophysical implications of our results and
provide simple fits to the exact expressions.

\vspace{0.7cm}
  
\noindent Pacs numbers: 13.10.+q, 13.15.+g, 14.70.Bh, 13.88.+e, 95.30.Cq.
\end{abstract}

\vspace{4.cm}

\leftline{} 
\leftline{} 
\leftline{CERN-TH/98-241}
\leftline{August 1998}  

\renewcommand{\baselinestretch}{1.2}

\newpage \section{Introduction}
Photon--neutrino processes are potentially of interest in 
astrophysics and cosmology. 
Four-leg elastic scattering processes are suppressed
by powers of $\omega/M_W$, where $\omega$ is the centre-of-mass
energy of the collision and $M_W$ the $W$ boson mass
\cite{Yang}. Their cross sections are therefore 
too small to be of astrophysical interest \cite{dic2}. 
On the other hand, five-leg processes involving two neutrinos 
and three photons, such as
\begin{eqnarray}
\gamma\nu    &\to& \gamma\gamma\nu     \nonumber \\
\gamma\gamma &\to& \gamma\nu\bar\nu    \nonumber \\
\nu\bar\nu   &\to& \gamma\gamma\gamma  \label{eq4}\,, 
\end{eqnarray}
are not negligible. In fact, the extra $\alpha$ in the cross section 
is compensated by an interchange of the $\omega/M_W$ suppression by an 
$\omega/m_e$ enhancement \cite{nous}.  

In ref. \cite{dic3}, Dicus and Repko derived an effective Lagrangian for the 
above five-leg photon--neutrino interactions, 
by substituting one photon with a neutrino current in the Euler--Heisenberg
Lagrangian that describes the photon--photon scattering \cite{EH}.
Some of the formulae reported in ref. \cite{dic3} were in disagreement 
with the results derived earlier by {\mbox Van~Hieu}
and Shabalin \cite{russian}. 
To settle this question, in a recent work \cite{nous} we computed
the first and the second process in \mbox{eq. (\ref{eq4})}, 
in the framework of the effective theory, 
confirming the results reported in ref. \cite{dic3}. 
We also justified the derivation of the five-leg effective vertex starting from the Euler--Heisenberg Lagrangian.

The effective approach gives reliable results for energies below 
the threshold for $e^+ e^-$ pair production, while its
extrapolation to energies above $1$ MeV, interesting to study, for example 
supernova dynamics, is suspect. Therefore, an exact calculation 
of the processes in \mbox{eq. (\ref{eq4})} is 
important in order to definitively assess their role in astrophysics
and the range of validity of the effective theory.
Such a calculation, assuming massless neutrinos, 
is the main ingredient of this paper.
The range of energy in which the above reactions are relevant
is well below the $W$ mass, so that we treated the neutrino--electron 
coupling as a four-Fermi interaction.

Very recently, parallel work in the same direction has been carried out
by Dicus, Kao and Repko \cite{dic1}, so that we a chance to compare our
numerical results, finding complete agreement between the two
independent calculations.

An additional problem is searching for simple approximations 
to the exact result. We tried the following two approaches:
\begin{itemize} 
\item we expanded the exact amplitude in powers of 
$\omega/m_e$, to derive the first correction term to the effective theory; 
\item we fitted the curves obtained with the complete calculation.
\end{itemize}
It turned out that, as expected in ref. \cite{dic1}, an expansion in inverse
powers of the electron mass, is not sufficient to extend the 
range of validity of the effective theory beyond $\omega= m_e$. One is therefore
forced to use the complete calculation or fits to it.

The outline of the paper is as follows.
In section II, we give the essential steps of the calculation,
collecting our final formulae in an appendix.
Numerical results are presented in section III, together with
the large $m_e$ expansion and the fits.
Finally, in section IV, we consider some of the possible astrophysical and
cosmological implications of our results.
\section{Computation of the amplitude}\label{reduc}

The leading Standard Model (SM) diagrams contributing to the processes in
\mbox{eq. (\ref{eq4})} 
are given in \mbox{fig. 1}, where permutations of the photon legs are
understood. Momenta and polarization vectors are denoted by
the sets $\{i\},~\{j\}$ and $\{k\}$ ($i,~j,~k = 1\,,2\,,3$), with
$\{1\} \equiv \{p_1,\,\epsilon^\alpha(p_1,\lambda)\}$,
$\{2\} \equiv \{p_2,\,\epsilon^\beta(p_2,\rho)\}$ and
$\{3\} \equiv \{p_3,\,\epsilon^\gamma(p_3,\sigma)\}$, where
$\lambda\,,\rho\,,\sigma= \pm $. All momenta are 
incoming  and $q$, flowing in the direction of the fermionic arrow,
is the virtual integration momentum.

\bfig(300,120)
\sof(-50,20)
\flin{40,0}{110,0}  \flin{110,0}{180,0}
\wlin{110,0}{110,20}\
\flin{110,20}{90,40}
\flin{90,40}{110,60}
\flin{110,60}{130,40}
\flin{130,40}{110,20}
\glin{90,40}{65,40}{3}
\glin{110,60}{110,85}{3}
\glin{130,40}{155,40}{3}
\Text(63,-5)[t]{$p_4~~~~\nu$}
\Text(155,-5)[t]{$\nu~~~~p_5$}
\Text(115,12)[l]{$Z$}
\Text(98,55)[r]{$q$}
\Text(115,70)[l]{$\gamma$}
\Text(77,36)[t]{$\gamma$}
\Text(143,36)[t]{$\gamma$}
\Text(60,40)[r] {$\{i\}$}
\Text(160,40)[l]{$\{k\}$}
\Text(110,90)[b]{$\{j\}$}
\Text(131,55)[r]{$e$}
\Text(150,87)[bl]{$(a)$}
\sof(140,20)
\flin{40,0}{80,0}  \wlin{80,0}{140,0} \flin{140,0}{180,0}
\flin{80,0}{90,20} \flin{90,20}{110,32}
\flin{110,32}{130,20} \flin{130,20}{140,0}
\glin{90,20}{67,37}{3}
\glin{110,32}{110,62}{3}
\glin{130,20}{153,37}{3}
\Text(63,-5)[t]{$p_4~~~~\nu$}
\Text(155,-5)[t]{$\nu~~~~p_5$}
\Text(110,4)[b]{$W$}
\Text(98,33)[r]{$q$} 
\Text(115,50)[l]{$\gamma$}
\Text(75,23)[t]{$\gamma$} 
\Text(145,23)[t]{$\gamma$}
\Text(65,45)[r] {$\{i\}$}
\Text(155,45)[l]{$\{k\}$}
\Text(110,70)[b]{$\{j\}$}
\Text(131,33)[r]{$e$}
\Text(150,87)[bl]{$(b)$}
\efig{FIG. 1. SM leading diagrams contributing to five-leg 
photon--neutrino processes. }

At low energies, far from the scale of confinement, the leading contribution
is given by diagrams involving only electrons in the fermionic loop. 
It is precisely the appearance of $m_{e}$ as a scale,
instead of $M_{W}$ (which is the scale governing the  
four-leg photon--neutrino reactions),
that makes such five-leg processes relevant already at energies of the 
order of a few MeV. 

By denoting with $A_{ijk}$ and $B_{ijk}$ the contributions coming 
from the diagrams $(a)$ and $(b)$ in \mbox{fig. 1}, the 
total amplitude $M$ reads
\bea\label{M}
M(\lambda, \rho, \sigma) 
&=& \left [(A_{123}+ A_{321})+(A_{132}+ A_{231})+(A_{213}+ A_{312})\right ] 
\nonumber \\ &+&
\left [(B_{123}+ B_{321})+(B_{132}+ B_{231})+(B_{213}+ B_{312})\right] 
\, , \eea
where, for example
\bea
A_{123} &=&
-\sum_{\tau=\pm} v_{e}^{\tau}\,\Gamma_{\mu}
\int d^nq 
{\rm \,Tr\,}\left[\gamma^{\mu} w^{\tau} {1 \over Q_{(23)}^{-}} 
\rlap/\epsilon(p_3,\sigma)
{1 \over Q_{2}^{-}} 
\rlap/\epsilon(p_2,\rho)
{1 \over Q_{0}^{-}} 
\rlap/\epsilon(p_1,\lambda)
{1 \over Q_{-1}^{-}} \right]\, 
\nonumber \\ 
A_{321} &=&
-\sum_{\tau=\pm} v_{e}^{\tau}\,\Gamma_{\mu} 
\int d^nq 
{\rm \,Tr\,}
\left[\gamma^{\mu} w^{\tau} {1   
\over
Q_{1}^{-}} \rlap/\epsilon(p_1,\lambda)  {1 \over Q_{0}^{-}} 
\rlap/ \epsilon(p_2,\rho) {1 \over
Q_{-2}^{-}}
\rlap/\epsilon(p_3,\sigma)\,
{1 \over Q_{-(23)}^{-}}\right]\,.
\eea
In the previous equations and throughout the calculation, we used the notations
\bea
& Q_{\pm i}^{\mp}= \rlap/Q_{\pm i} \mp m_{e}\,, &~~~Q_{\pm (ij)}^{\mp}= \rlap/Q_{\pm (ij)} \mp m_{e}\,, \nonumber \\
& Q_{\pm i}= Q_0 \pm p_i\,,~~~                  &~~~Q_{\pm (ij)}= Q_0 \pm p_i \pm p_j\,,~~~Q_0= q\,, \nonumber \\
& D_{\pm i}= Q_{\pm i}^+\cdot Q_{\pm i}^-\,,    &~~~D_{\pm (ij)}= Q_{\pm (ij)}^+\cdot Q_{\pm (ij)}^-\,.
\eea
Furthermore, $w^{\pm}=(1 \pm \gamma_{5})/2$, and a dimensional 
regularization is used to compute each
separately divergent diagram. Finally, the current $\Gamma_\mu$ reads
\bea
\Gamma_\mu &=& {(g s_{W})}^3 {\left( {g \over 2 c_{W}} \right)}^2 \!
\left({1\over \Delta_{Z}} \right)
{1 \over {(2 \pi)}^{4}}\,{\bar v}_{+}(5) \gamma_{\mu}u_{-}(4)\,,
\eea
and
\bea 
\Delta_{Z}= {(p_{4}+p_{5})}^{2}-M_{Z}^{2}\,,~~~ 
v_{e}^{\pm}=v_{e}\pm a_{e}\,,~~~v_{e}=-\frac 1 2+2 s_{W}^{2}\,,~~~
a_{e}=\frac 1 2\,,
\eea
where $s_W$ and $c_W$ are the sine and the cosine of the Weinberg angle,
respectively. 
The reason why the terms in \mbox{eq. (\ref{M})} are collected in pairs
is because, when adding the two terms in each pair, 
and using the reversing invariance of the $\gamma$-matrix traces,
the $\gamma_5$ contribution cancels \cite{nous}. 
For example
\bea \label{as1}
A_{123}+A_{321}=-2 v_{e}\Gamma_{\mu} \int d^nq
{\rm \,Tr\,}\left[\gamma^{\mu} {1 \over
Q_{(23)}^{-}}\rlap/\epsilon(p_3,\sigma){1 \over Q_{2}^{-}} 
\rlap/\epsilon(p_2,\rho) {1 \over
Q_{0}^{-}}
\rlap/\epsilon(p_1,\lambda){1 \over Q_{-1}^{-}} \right]\,.
\eea
Exactly the same work can be applied to each pair of diagrams
of type $B$. For instance, we get 
\bea\label{bs}
&& B_{123}+B_{321}=\,-2\, \Delta_Z\, c_W^2\, \Gamma_{\mu} \int d^nq
\,{\rm Tr\,}\left[\gamma^{\mu} {1
\over
Q_{(23)}^{-}} \rlap/\epsilon(p_3,\sigma) {1 \over Q_{2}^{-}}
\rlap/\epsilon(p_2,\rho) {1 \over
Q_{0}^{-}}
\rlap/\epsilon(p_1,\lambda) {1 \over Q_{-1}^{-}} \right]\,
\frac{1}{\Delta_{W}(q)}\,, \nonumber \\
&& ~~
\eea
where $\Delta_{W}(q)\equiv (q+p_{2}+p_{3}+p_{5})^{2}-M_{W}^{2}$.
Since we are interested in the large-$M_W$ (or equivalently 
low-energy) limit,  
we are allowed to substitute \cite{nous}
\bea
\Delta_{Z} \sim -M_{Z}^{2}~~{\rm and}~~\Delta_{W} \sim -M_{W}^{2}\,, 
\eea
so that the sum of the diagrams in \mbox{eqs. (\ref{as1}) and (\ref{bs})} gives
\bea \label{set1}
&& A_{123}+A_{321}+B_{123}+B_{321} =
-2 (v_{e}+1) \Gamma_{\mu} I_{1}^{ \mu}(p_1, p_2, p_3, \lambda,
\rho, \sigma)~~~~~~{\rm with} \nonumber \\
&& I_{1}^{\mu}(p_1, p_2, p_3, \lambda,\rho, \sigma) = 
\int d^nq
{\rm \,Tr\,}\left[\gamma^{\mu} {1 \over Q_{(23)}^{-}} 
\rlap/ \epsilon (p_3,\sigma)
{1 \over Q_{2}^{-}} 
\rlap/ \epsilon (p_2,\rho)
{1 \over Q_{0}^{-}}
\rlap/ \epsilon (p_1,\lambda)
{1 \over Q_{-1}^{-}} \right]\,, \label{master}
\eea
where the replacement $\Delta_{Z} \sim -M_{Z}^{2}$ has to be
performed also in the current $\Gamma_\mu$.

Therefore, the total amplitude reads
\bea \label{totamp}
M(\lambda, \rho, \sigma)= -2 (1+v_{e})\, \Gamma_{\mu}\, 
     [I_{1}^{ \mu}(p_1, p_2, p_3, \lambda,\rho, \sigma)
     &+&I_{2}^{ \mu}(p_1, p_2, p_3, \lambda,\rho, \sigma) \nonumber \\
     &+&I_{3}^{ \mu}(p_1, p_2, p_3, \lambda,\rho, \sigma)]\,,
\eea
where $I_{2}^\mu$ and $I_{3}^\mu$ come from the remaining pairs 
of terms in \mbox{eq. (\ref{M})} and can be obtained from $I_{1}^{\mu}$ as follows
 \bea \label{perm}
&&I_{2}^\mu: \qquad \qquad \qquad p_{2} \leftrightarrow
p_{3}\,, \quad \epsilon(p_2,\rho) \leftrightarrow 
\epsilon(p_3,\sigma) \nn \\
&&I_{3}^\mu: \qquad \qquad \qquad p_{2} \leftrightarrow p_{1}\,,
\quad  \epsilon(p_2,\rho) \leftrightarrow \epsilon(p_1,\lambda)\,.   
\eea
The reduction of $M(\lambda, \rho, \sigma)$ to scalar one-loop
integrals can then be performed with the help 
of the technique described in ref. \cite{Pittau}. 
The general philosophy of such a method is using the $\gamma$ algebra 
in the traces to reconstruct the denominators appearing in the loop integrals,
rather than making a more standard tensorial decomposition \cite{Passarino}.
The algorithm can be iterated in such a way that only scalar and
rank-one functions appear at the end of the reduction, at worst
together with higher-rank two-point tensors.

In this paper we content ourselves with giving the first step 
of this reduction, namely all results expressed only in terms of
\begin{itemize}
\item scalar functions with 3 and 4 denominators, 
\item rank-1 integrals with 3 and 4 denominators,
\item rank-2 integrals with 3 denominators,   
\item rank-3 functions with 3 denominators.
\end{itemize}
This already provides an important simplification with respect
to the standard decomposition, in that the computation of tensors such as
\bea
T^{\mu \nu;\, \mu \nu \rho;\, \mu \nu \rho \sigma} =
\int d^nq \,{q^\mu q^\nu ;\, q^\mu q^\nu q^\rho;\, q^\mu q^\nu q^\rho q^\sigma  \over
D_0\,D_{-1}\,D_2\,D_{(23)}}
\eea 
is completely avoided.

The key ingredient in the case at hand is a suitable choice of the
polarization vectors \cite{kleiss}
\bea \label{polari}
\rlap/\epsilon(p_1,\lambda) &=& N\,
  \left[           \rlap/p_2 \rlap/p_3 \rlap/p_1 w_\lambda +
         w_\lambda \rlap/p_1 \rlap/p_3 \rlap/p_2  \right] \nonumber \\
\rlap/\epsilon(p_2,\rho) &=& N\,
  \left[           \rlap/p_1 \rlap/p_3 \rlap/p_2 w_\rho +
         w_\rho    \rlap/p_2 \rlap/p_3 \rlap/p_1  \right] \nonumber \\
\rlap/\epsilon(p_3,\sigma) &=& N\,
  \left[           \rlap/p_1 \rlap/p_2 \rlap/p_3 w_\sigma +
         w_\sigma  \rlap/p_3 \rlap/p_2 \rlap/p_1  \right] \,,
\eea
where
\bea\label{nfac}
N  =  \left(\frac{1}{4\,(p_1\cdot p_2)(p_1\cdot p_3)(p_2\cdot
  p_3)}\right)^{\frac{1}{2}}~~{\rm and}
~~w_\pm =  \frac{1}{2}(1 \pm \gamma_5)\,.
\eea
Inserting \mbox{eq. (\ref{polari})} in the expression for $I_{2}^\mu$ and 
using the identity
\bea
  \rlap/ p_2\rlap/ p_3\rlap/ p_1 w^\lambda + w^\lambda \rlap/ p_1\rlap/ p_3\rlap/ p_2=
-(\rlap/ p_3\rlap/ p_2\rlap/ p_1 w^\lambda + w^\lambda \rlap/ p_1\rlap/ p_2\rlap/ p_3)
+2\,(p_2 \cdot p_3)\, [Q_0^{-}-Q_{-1}^{-}]\,,
\eea
gives 
\bea\label{i2exp}
 I_{2}^\mu(p_1,p_2,p_3,\lambda,\rho,\sigma) &=&
-I_{1}^\mu(p_1,p_3,p_2,\lambda,\sigma,\rho)+ J_{2}^\mu(p_1,p_2,p_3,\rho,\sigma)~~~~~{\rm with} \nonumber\\
J_{2}^\mu(p_1,p_2,p_3,\rho,\sigma)&=& 2\,(p_2 \cdot p_3)\,N^3\, \left\{ 
\int d^nq
{\rm \,Tr\,}\left[\gamma^{\mu} 
{1 \over Q_{(23)}^{-}}(\rlap/ p_1\rlap/ p_3\rlap/ p_2 w^\rho + w^\rho \rlap/ p_2\rlap/ p_3\rlap/ p_1)
\right. \right. \nonumber \\ 
&\times& \left. \left.
{1 \over Q_{3}^{-}} (\rlap/ p_1\rlap/ p_2\rlap/ p_3 w^\sigma + w^\sigma \rlap/ p_3\rlap/ p_2\rlap/ p_1) 
\left({1 \over Q_{-1}^{-}}-{1 \over Q_{0}^{-}}\right)\right] \right\}\,.
\eea
With similar arguments we also get
\bea\label{i3exp}
 I_{3}^\mu(p_1,p_2,p_3,\lambda,\rho,\sigma) &=&
-I_{1}^\mu(p_2,p_1,p_3,\rho,\lambda,\sigma)+ J_{3}^\mu(p_1,p_2,p_3,\lambda,\rho) ~~~~~{\rm with} \nonumber\\
J_{3}^\mu(p_1,p_2,p_3,\lambda,\rho)&=& 2\,(p_1 \cdot p_2)\,N^3\, \left\{ 
\int d^nq
{\rm \,Tr\,}\left[\gamma^{\mu}\left({1 \over Q_{1}^{-}}-{1 \over Q_{(13)}^{-}}\right) 
(\rlap/ p_2\rlap/ p_3\rlap/ p_1 w^\lambda + w^\lambda \rlap/ p_1\rlap/ p_3\rlap/ p_2)
\right. \right. \nonumber \\ 
&\times& \left. \left.
{1 \over Q_{0}^{-}} (\rlap/ p_1\rlap/ p_3\rlap/ p_2 w^\rho + w^\rho \rlap/ p_2\rlap/ p_3\rlap/ p_1) 
{1 \over Q_{-2}^{-}}
\right] \right\}\,.
\eea
Therefore, since $J_{2}^\mu$ and  $J_{3}^\mu$ are directly expressed in terms of differences of three-point functions,
$I_{1}^\mu$ given in \mbox{eq. (\ref{master})} is the only master integral we need to decompose in terms of
simpler tensorial structures.
By inserting \mbox{eq. (\ref{polari})} into \mbox{eq. (\ref{master})}
for each combination of photon helicities, 
we obtain four different expressions.
For example, when $\rho=\sigma= -\lambda$:
\bea 
 I_{1}^\mu(p_1,p_2,p_3,\lambda,-\lambda,-\lambda) &=& N^3
\int d^nq \frac{1}{D_0\,D_{-1}\,D_{2}\,D_{(23)}} A^\mu(\lambda,-\lambda,-\lambda)\,,
\eea
with
\bea\label{exam}
A^\mu(\lambda,-\lambda,-\lambda) &=& 
m_e^2\, ([\mu Q_{(23)}123231 Q_0 231]_\lambda
    +[\mu Q_{-1}132 Q_0 132321  ]_\lambda \nonumber \\
  &+&[\mu 231 Q_0 231 Q_2 321]_\lambda
    +[\mu 321 Q_{2} 231 Q_0 231]_\lambda) \nonumber \\
&+&[\mu Q_{-1}  132 Q_0 132 Q_2 123 Q_{(23)}]_\lambda 
  +[\mu Q_{(23)}123 Q_2 132 Q_0 132 Q_{-1}]_\lambda\,,
\eea
and where we used the notation $[\mu i j k \cdots ]_\lambda= 
{\rm Tr} [\gamma^\mu \rlap/p_i \rlap/p_j \rlap/p_k \cdots w^\lambda]$.

To illustrate how the reduction works, take for example the last term
in \mbox{eq. (\ref{exam})}.
Since $p_3 \cdot p_3 = 0$, we may rewrite
\bea
         [\mu Q_{(23)}123 Q_2 132 Q_0 132 Q_{-1}]_\lambda=
        -[\mu Q_{(23)}123 1 Q_232 Q_0 132 Q_{-1}]_\lambda \,,
\eea
and, from the identity
\bea
\rlap/Q_{2}\rlap/p_3\rlap/p_2\rlap/Q_0= 
 (D_{(23)}-D_{2})\rlap/p_2\rlap/Q_0 
+(D_{0}-D_{2})\rlap/p_3\rlap/Q_0 
+(D_{0}+m_e^2)\rlap/p_3\rlap/p_2\,, 
\eea
the first step of the denominator reconstruction immediately follows.
As a second example, consider a term such as $[\mu \cdots 1 Q_0 1 \cdots ]_\lambda$.
Since $ \rlap/ p_1 \rlap/ q \rlap/ p_1= \rlap/ p_1 (D_0 - D_{-1})$, we immediately get
\bea
[\mu \cdots 1 Q_0 1 \cdots ]_\lambda= (D_0 - D_{-1})\,[\mu \cdots 1 \cdots ]_\lambda\,,~~~~~{\rm etc}\,.
\eea
The final result of such a procedure is given in the appendix.
An important remark is in order here. To obtain compact expressions, 
we made a large use of the Kahane--Chisholm manipulations over 
$\gamma$ matrices \cite{kahane}. Such identities
are strictly four-dimensional, while we are, at the same time,  
using dimensional regularization.
Our solution is splitting, {\em before any trace manipulation}, 
the $n$-dimensional integration momentum appearing in the traces as 
\cite{Pittau}
\bea\label{splitq}
q\to q +{\tilde{q}}\,,
\eea
where $q$ and ${\tilde{q}}$ are the four-dimensional and 
$\epsilon$-dimensional components ($\epsilon= n-4$), respectively, 
so that $q \cdot \tilde{q}= 0$.
The $\gamma$ algebra can then be safely performed in four dimensions, at the price of having additional terms.
In fact, the splitting in \mbox{eq. (\ref{splitq})} is equivalent to redefining $m_e^2\to m_e^2- {\tilde{q}}^2$ from
the beginning. The net effect is then the appearance of extra integrals containing powers of $\tilde{q}^2$ in the numerator, 
whenever $m_e^2$ is present in the formulae reported in the appendix.
The computation of such integrals in the limit $\epsilon \to 0 $ is 
straightforward \cite{Pittau}.
For example 
\bea 
&& \int d^nq {{\tilde{q}}^4\over D_0 \ D_{-1}\ D_2\ D_{(23)}}= -i {\pi^2\over 6}+{\cal{O}}(\epsilon)\,, \nonumber \\
&& \int d^nq { q_\mu {\tilde{q}}^2\over
 D_0 \ D_{-1}\ D_2}= i {\pi^2\over 6}\,(p_2-p_1)_\mu+{\cal{O}}(\epsilon)\,.
\eea 

A standard Passarino--Veltman decomposition \cite{Passarino}
of the simple remaining tensorial structures in terms of scalar 
loop functions, concludes our calculation. We implemented the
outcoming formulae in a Fortran code, performing the phase-space
integration by Monte Carlo. Numerical results are reported 
in the next section.

Our formulae remain valid also when including all neutrino species.
In this case, only the first diagram in \mbox{fig. 1$a$} contributes, at 
leading order in ${\omega}/{m_e}$,
because the second one is suppressed by powers of $\omega/m_{\mu,\tau}$. Therefore, the
inclusion of all neutrinos can be achieved by simply replacing $(1+v_e)$ with $(1+ 3\,v_e)$ in \mbox{eq. (\ref{totamp})}.
However, we only considered $\nu_e$ in our numerical results.
\section{Results}\label{plots}

In this section, we give numerical results for the three reactions in
\mbox{eq. (\ref{eq4})} and present 
simple approximations to the full computation. Our main motivations are
to assess the range of validity of the effective theory, and go beyond it.

The total cross sections computed using the effective Euler--Heisenberg
Lagrangian are \cite{nous,dic3} 
\bea
\label{apsig}
&&\sigma^{eff}(\gamma\nu\to\gamma\gamma\nu)= 
  \frac{262}{127575} ~k^2~\left(\frac{\omega}{m_e}\right)^{10} \nonumber \\ 
&&\sigma^{eff}(\gamma\gamma\to\gamma\nu\bar\nu)=
  \frac{2144}{637875}~k^2~\left(\frac{\omega}{m_e}\right)^{10} \nonumber \\
&&\sigma^{eff}(\nu\bar\nu\to\gamma\gamma\gamma)= 
  \frac{136}{91125}  ~k^2~\left(\frac{\omega}{m_e}\right)^{10}\,, 
\eea
with $k= G_F\,m_e\,(1+v_e)\,\alpha^{3/2}/\pi^2$. 
Effective and exact computations are compared in 
\mbox{figs. \ref{complet1}, \ref{complet2} and \ref{complet3}}, 
for the three cases.
Furthermore, table I shows the ratio between exact cross section and $\sigma^{eff}$ for several
values of $\omega/m_e$.  

From the above figures and numbers it is clear that, for all three processes, 
the effective theory is valid only when, roughly, $\omega/m_e \leq 2$, 
as expected. At larger $\omega$, the exact
computation predicts a softer energy dependence with respect to the 
$\left({\omega}/{m_e}\right)^{10}$ behaviour given by the effective
Lagrangian. 

All above results are in full agreement with those reported in 
ref. \cite{dic1}.
Notice also that exact and effective predictions approach each other
in the limit of vanishing $\omega$, as it should be. This provides
us with an additional, strong numerical check on the correctness of our 
computation.

Since the exact formulae are quite involved, we found it convenient to look for
an approximation above the point $\omega/m_e = 2$.
With this aim, we performed a large-$m_e$ expansion 
of the integrals in \mbox{eqs. (\ref{master}) and (\ref{perm})}, 
to determine the 
${\cal O}(1/m_e^6)$ correction 
to the effective amplitude
for $\gamma\nu\to\gamma\gamma\nu$.
By iteratively applying the equation
\bea 
{1 \over ({(q+k)}^{2}-m_{e}^{2})}={1 \over {q}^{2}-m_{e}^{2}} -
{k^{2} + 2 q \cdot k \over (q^{2}-m_{e}^{2}) ({(q+k)}^{2}
 -m_{e}^{2})}\,,
\eea
and systematically discarding terms smaller than 
${\cal O}(1/m_e^6)$, one is left with only contributions of the kind
\bea\label{mexp}
 m_{e}^{r}\, p^{s}\, \int d^nq   {q^{t} \over {(q^2-m_{e}^2)}^l}\,, 
\eea
in which the external momentum dependence in factorized out from the integrals.
In the above formula, $p^s$ stands for any product 
of $s$ external momenta, and $t$, $l$, $r$, $s$ 
are constrained by $4+r+t-2l+s=0$. Thus, when $r+s=0$ the integral in
\mbox{eq. (\ref{mexp})} is divergent, when $r+s=2$ it is finite and proportional to
$1/m_{e}^{2}$, and so on.

By summing all relevant contributions, we found 
\bea\sigma(\gamma\nu\to\gamma\gamma\nu)={m_e^2 G_F^2 (1+v_e)^2 \alpha^3\over
127575\  \pi^4 }
\left(262\left({\omega\over m_e}\right)^{10} -{3163\over 20}\left({\omega\over
m_e}\right)^{12}\right)\label{tot}\,.
\eea
The first term in the r.h.s. of \mbox{eq. (\ref{tot})}
coincides with the result given in \mbox{eq. (\ref{apsig})}, so this is an extra check of our
computation. The second term has the right sign but, as expected 
in ref. \cite{dic1}, it does not give 
numerical predictions that are useful to extend the effective 
theory beyond $\omega= m_e$. The only option is then
to fit the curves in \mbox{figs.
\ref{complet1}, \ref{complet2} and \ref{complet3}}.
The results of the fits are

\bea\label{fit1}
&&\sigma(\gamma\nu\to\gamma\gamma\nu)=
\sigma^{eff}(\gamma\nu\to\gamma\gamma\nu)\times 
 r^{-2.76046}\nn \\ &&~~~\times
{\rm exp}\,[2.13317-2.12629 \, {\rm log}^{2}(r)+0.406718 \, {\rm log}^{3}(r)  
-0.029852\, {\rm log}^{4}(r)]\,,  \nonumber \\
&&\sigma(\gamma\gamma \to \gamma\nu {\bar \nu})=
\sigma^{eff}(\gamma\gamma \to \gamma\nu {\bar \nu}) \times
r^{-7.85491}\nn\\&&~~~\times
{\rm exp}\,[4.42122+0.343516\, {\rm log}^{2}(r)-0.114058 \, {\rm log}^{3}(r)  
+0.0103219\,{\rm log}^{4}(r)]\,, \nonumber \\
&&\sigma(\nu {\bar \nu}\to \gamma \gamma \gamma)=
\sigma^{eff}(\nu {\bar \nu}\to \gamma \gamma \gamma)\times
 r^{-6.57374}\nn\\&&~~~\times
{\rm exp}\,[5.27548-0.689808 \, {\rm log}^{2}(r)+0.15014 \, {\rm  log}^{3}(r) 
-0.0123385 \, {\rm log}^{4}(r)]\,,
\eea
where the effective cross sections $\sigma^{eff}$ are given in 
\mbox{eq. (\ref{apsig})}, and $r= \omega/m_{e}$.
All the above fits are valid in the energy range $1.7 < r < 100 $.
\section{Discussion and concluding remarks}\label{conclusion}
The processes $\nu\gamma\to\nu\gamma\gamma$, 
$\nu\bar\nu\to\gamma\gamma\gamma$ and $\gamma\gamma\to\gamma\nu\bar\nu$ 
are of potential interest in stellar evolution and cosmology.
The first two reactions can affect the mean free path of neutrinos
inside the supernova core, while
the last one is a possible cooling mechanism for hot objects \cite{teplitz}.
The relevance of such processes depends on the
size of the cross sections when varying the centre-of-mass energy $\omega$, and may be definitively
assessed only through complete and detailed Monte Carlo simulations. Nevertheless, 
some simple considerations can be made, also in connection 
with speculations that recently appeared in the literature.

Basing their results on the assumption
\bea\label{steplitz}
    \sigma(\gamma\nu\to\gamma\gamma\nu)= \sigma_0 \left(\frac{\omega}{1~{\rm MeV}} \right)^\gamma\,,
~~~~\sigma_0= 10^{-52}\,{\rm cm}^2\,,
\eea
and on the data collected from supernova 1987A, the authors of ref. \cite{teplitz} 
fitted the exponent $\gamma$ in 
\mbox{eq. (\ref{steplitz})} to be less than $8.4$, for $\omega$ of the order
of a few MeV. The physical requirement behind this is that neutrinos 
of a few MeV should immediately leave the supernova, 
so that their mean free path is constrained to be larger than $10^{11}$ cm.

The effective theory predicts $\gamma= 10$, while, using the exact calculation, a softer energy dependence is
observed in the region of interest (see \mbox{fig. \ref{complet1}}). 
A fit to the exact curve 
gives $\gamma \sim 3$ for $1~{\rm MeV} < \omega < 10~{\rm MeV}$,
thus confirming the expectations of ref. \cite{teplitz}.

A second interesting quantity is the range of parameters for which the neutrino mean free path for such reactions
is inside the supernova core, therefore affecting its dynamics.
Always in ref. \cite{teplitz} it was found, with 
the help of Monte Carlo simulations, that for several choices 
of temperature and
chemical potential, and assuming the validity of the effective theory ($\gamma= 10$), this happens when
$\omega \ge 5$ MeV. Since the exact results are now available, it would be of extreme interest 
to see how the above prediction is affected. 
More in general, we think that the reactions
in \mbox{eq. (\ref{eq4})} should be included in 
supernova codes.

In ref. \cite{dic3}, it was speculated that these processes  
could also have some relevance in cosmology.  
Consider, in fact, the mean number $\bar N$ of neutrino collisions,
via the first reaction in \mbox{eq. (\ref{eq4})},
in an expansion time $t$ equal to the age of the Universe \cite{Peebles}: 
\bea
\bar N=\sigma(\gamma\nu\to\gamma\gamma\nu) n_\nu c t\,,~~~~~~
n_\nu={\rm neutrino~number~density}\,.
\eea 
In the above formula, $n_\nu$ and $t$ can be rewritten in terms of the photon energy at thermal equilibrium
($\omega \sim k T$), expressed in units of $10^{10}$ K. By denoting this quantity by $T_{10}$, we get
\bea
n_\nu=1.6 \times 10^{31} T_{10}^3\ {\rm cm}^{-3}\,, 
~~~t=2 \ T_{10}^{-2}\,\, {\rm s}.
\eea
When $\bar N$ is large, the neutrinos are in thermal contact with matter and radiation, while, for 
$\bar N\sim 1$ (namely $\sigma\sim 10^{-42}
T_{10}^{-1}\,\,\mathrm{cm}^{2}$), 
the neutrinos decouple. 
By using the formula in \mbox{eq. (\ref{apsig})}, the resulting decoupling 
temperature
is $T_{10}\ \sim\ 9.5$, namely $\omega\ \sim\ 8.2$ MeV, therefore outside the range of validity
of the effective theory. By repeating the same analysis with the exact result, we found instead
that $\bar N$ becomes of the order of 1 at $\omega\ \sim\ 1$ GeV.
At these energies, other processes enter the game; for instance reactions involving different leptons,
quarks and hadrons inside the loop in \mbox{fig. 1$a$}. Therefore, the reaction $\gamma\nu\to\gamma\gamma\nu$ is no longer the
only relevant process. In conclusion, the five-leg reactions 
in \mbox{eq. (\ref{eq4})} are unlikely to be important for a study of the neutrino
decoupling temperature, contrary 
to what the effective theory seemed to suggest.

\section*{Acknowledgements}
We thank the authors of ref. \cite{dic1} for having informed us about 
their recent computation. 
We also wish to thank M. B. Gavela, G. F. Giudice and O. P\`ene 
for helpful discussions, and G. Passarino for numerical checks on the loop
functions. 
J.M. acknowledges the financial support from a
Marie Curie EC Grant (TMR-ERBFMBICT 972147).

\clearpage
\appendix 
\begin{center}{\bf Appendix} 
 \end{center}
 
\noindent
We list here the result of the decomposition
of the master integral in \mbox{eq. (\ref{master})}, performed with the technique of
ref. \cite{Pittau}.
The starting point is the formula
\bea
 I_{1}^\mu(p_1,p_2,p_3,\lambda,\rho,\sigma) &=& N^3
\int d^nq \frac{1}{D_0\,D_{-1}\,D_{2}\,D_{(23)}} 
A^\mu(\lambda,\rho,\sigma)\,,
\eea
with $N$ given in \mbox{eq. (\ref{nfac})}.

By defining $(i \cdot j) \equiv p_i \cdot p_j$ and 
$m \equiv m_e$, the four possible helicity configurations read
\begin{itemize}
\item $\lambda= -\rho=-\sigma$:
\end{itemize}
\bea \label{apmm2}
A^\mu&&(\lambda,-\lambda,-\lambda) = 
-4\,(1 \cdot 2) (2\cdot 3)\,D_0\,\left\{ [\mu Q_0 132 Q_0 1 3]_\lambda
     +2\,(1 \cdot 3)[\mu Q_{(23)} 132 Q_{-1}]_\lambda \right\}
\nonumber \\ &&
+(D_{(23)}-D_{2}) \left\{
   4\,m^2\,(1 \cdot 2)\,\left[\; 
             [\mu 231Q_0231]_\lambda
          - 2[\mu 1(Q_02-2Q_0)]_\lambda (1\cdot 3)(2\cdot 3)\;
                        \right]\right.
\nonumber \\ &&
           \left. - [\mu Q_{(23)}1231Q_02312Q_{-1}]_\lambda 
            - [\mu Q_{0}1321Q_0231Q_02]_\lambda \ \right \}
\nonumber \\ &&
+(D_0-D_2) \left\{
          2\,( 1\cdot  3)\,m^2\, \left[\; 
          4\,( 1 \cdot 2) (2 \cdot 3) [\mu 1(3Q_0-Q_03)]_\lambda  
                                  + [\mu (32-23) Q_01231]_\lambda 
                           \; \right]\right.
\nonumber \\ &&\left.
          - 2\,[\mu Q_{(23)}123Q_0132Q_{-1}]_\lambda (1\cdot 3)
          +  [\mu Q_01321Q_023Q_013]_\lambda \right\}
\nonumber \\ &&
-16\,m^2\,(1\cdot 2)(1\cdot 3)(2\cdot 3)(D_0-D_{-1})[\mu 32Q_0]_\lambda
\nonumber \\ &&
         +8\,(1\cdot 2)(1\cdot 3)(2\cdot 3)(D_{0}+m^{2})\,m^2\, \left\{
[\mu 132]_\lambda-[\mu 123]_\lambda+[\mu 321]_\lambda-[\mu 231]_\lambda\right\}
\nonumber \\ &&
+8\, m^2 (1\cdot 2)(1\cdot 3)(2\cdot 3) \left\{
-[\mu Q_01231]_\lambda
+2\,(2\cdot 3)[\mu 1(Q_02-2Q_0)]_\lambda\right.
\nonumber \\ &&\left.
-2\,(1\cdot 2)[\mu 231]_\lambda
+[\mu 2312Q_0]_\lambda
-2\,(1\cdot 3)[\mu 32Q_{-1}]_\lambda \right\}
\nonumber \\ &&
-4\, m^2 (1\cdot 3)(2\cdot 3)[\mu 321Q_0231]_\lambda\,.
\eea
\begin{itemize}
\item $\lambda= \rho= \sigma$:
\end{itemize}

\noindent\bea
A^\mu&&(\lambda,\lambda,\lambda) = 8\, m^2 (1\cdot 2)(1\cdot 3)(2\cdot 3)
(D_{(23)}-D_{2})[\mu 2(Q_01-1Q_0)]_\lambda \nonumber \\ &&
-4\,( 1 \cdot 3) (2 \cdot 3)(D_0-D_2) 
\left\{[\mu Q_0 1Q_2 3 2 1 Q_{(23)}]_\lambda\right.
\nonumber\\ &&
\left. + m^2\, \left[\;[\mu 1 2 3  Q_2 1]_\lambda- [\mu 1 3 2  1 Q_0 ]_\lambda
 +2\,( 1 \cdot 3)  [\mu 2 1 Q_0]_\lambda  - 2\,( 1 \cdot 2)  
[\mu 3 Q_0 1]_\lambda
 \;\right]\right\}
\nonumber \\ &&
+4\,( 1 \cdot 3) (2 \cdot 3)(D_0-D_{-1}) 
\left\{[\mu Q_2 3 2 1  Q_0 2 Q_{-1}]_\lambda
\right.
\nonumber\\ &&
\left. + m^2 \left[\;[\mu 2 3 1 2 Q_0]_\lambda +  [\mu 2   Q_0 1 2 3]_\lambda
 +2\,( 2 \cdot 3)\,([\mu 2 1 Q_{(23)}]_\lambda - [\mu 121]_\lambda)  
- 2\,( 1 \cdot 2) 
  [\mu Q_0  32]_\lambda
 \;\right]\right\}
\nonumber \\ &&
         +16\,(1\cdot 2)(1\cdot 3)(2\cdot 3)(D_{0}+m^{2}) m^2 \left\{
\left[\;3\,(2\cdot 3)-2\,(1\cdot 2)\;\right]\,[\mu 1]_\lambda \right.
\nonumber\\ &&\left.- 2\,[\mu 123]_\lambda -2\,\left[\;(1\cdot 3)+(1\cdot
2)\;\right]\,[\mu 2]_\lambda \right\}
\nonumber \\ &&
-8\, m^2 (1\cdot 2)(1\cdot 3)(2\cdot 3) \left\{
[\mu 21Q_032]_\lambda-[\mu 123Q_01]_\lambda
+2\,(2\cdot 3) \{\;[\;m^2-4\,(1\cdot 2)\;]\,[\mu 1]_\lambda\right.
\nonumber \\ &&\left.
+[\mu Q_012]_\lambda+[\mu Q_013]_\lambda +[\mu 12 Q_0]_\lambda
-[\mu 123]_\lambda \}\;  \right\}
+4\, m^2 (1\cdot 3)(2\cdot 3)[\mu 21Q_03123]_\lambda\,.
\eea
\begin{itemize}
\item $\lambda= \rho= -\sigma$:
\end{itemize}

\noindent\bea
A^\mu&&(\lambda,\lambda,-\lambda) = -16\, m^2 (1\cdot 2)(1\cdot 3)(2\cdot 3)
(D_{(23)}-D_{2})[\mu 12Q_0]_\lambda \nonumber \\ &&
-4\,( 1 \cdot 3) (2 \cdot 3)(D_0-D_2) \left\{ 
[\mu Q_{0}1 Q_2 123Q_2]_\lambda\right.
\nonumber\\ &&
\left. + m^2 \left[\; [\mu 3 2 1  Q_2 1]_\lambda
- [\mu 2 1 3 2 1 ]_\lambda- [\mu 3 1 3 2 1 ]_\lambda
 +4( 1\cdot 2)  [\mu 1 3 Q_0]_\lambda  
\;\right]\right\}
\nonumber \\ &&
+4\,( 1 \cdot 3) (2 \cdot 3)(D_0-D_{-1}) \left\{[\mu Q_{(23)}1 2 3  Q_0 2 Q_{-1}]_\lambda
\right.
\nonumber\\ &&
\left. + m^2 \left[\:[\mu 2 3  Q_0 1 2 ]_\lambda   
 +2\,( 1 \cdot 2)  \{\;[\mu 2 3 Q_{2}]_\lambda - [\mu 321]_\lambda 
 - 4\,[\mu 2 3 Q_0]_\lambda  + 2\,( 2 \cdot 3) 
  [\mu Q_0 ]_\lambda\;\}\;
 \right]\right\}
\nonumber \\ &&
         +16\,(1\cdot 2)(1\cdot 3)(2\cdot 3)(D_{0}+m^{2}) m^2 \left\{
[\mu 1 2 3]_\lambda-(2\cdot 3)[\mu 1]_\lambda \right\}
\nonumber \\ &&
+8\, m^2 (1\cdot 2)(1\cdot 3)(2\cdot 3) \left\{[\mu 23Q_012]_\lambda
- [\mu 2312Q_0]_\lambda+[\mu 321Q_01]_\lambda-[\mu Q_01321]_\lambda
\right.
\nonumber \\ &&   + m^2\,[\mu (23-32)1]_\lambda +2 (1\cdot 3 )
[\mu 32Q_{-1}]_\lambda  \nonumber \\ &&
 +2\,(2\cdot 3 )\left[\;
-[\mu Q_012]_\lambda +2\,[\mu 12Q_0]_\lambda 
-2\,(1\cdot 2 )[\mu 2]_\lambda\;\right]
\nonumber \\ &&\left. 
 +2\,(1\cdot 2 )\left[\;
-2\,[\mu23 Q_0]_\lambda - [\mu 32Q_0]_\lambda  
+2\,(2\cdot 3)[\mu 1]_\lambda\;\right]
 \right\}
\nonumber \\ &&
+4\, m^2 (1\cdot 3)(2\cdot 3)\left\{[\mu 3213Q_021]_\lambda -2\,(2\cdot 3 )
[\mu 21Q_012]_\lambda \right\}\,.
\eea
\begin{itemize}
\item $\lambda= -\rho= \sigma$:
\end{itemize}
\noindent\bea
A^\mu&&(\lambda,-\lambda,\lambda)=
  4\,D_0 (1 \cdot 2) (2 \cdot 3) \left\{2\,(1 \cdot 3) 
     \{ [\;m^2 + 2\,(1 \cdot 2)\;]\, [\mu 2 3 1]_\lambda 
    \right.
    \nonumber \\ && \left.
       + 2\,[\;(1 \cdot 2) [\mu 3 2 Q_0]_\lambda 
+ (2 \cdot 3) [\mu Q_0 1 2]_\lambda + 
           (2 \cdot 3) [\mu Q_0 1 3]_\lambda\;]\} + 
           [\mu Q_0 1 3 2 Q_0 3 1]_\lambda \right\}
 \nonumber \\ &&
 + 4\,(D_2 - D_{(23)}) (1 \cdot 2) (2 \cdot 3) 
   \left\{4\,(1 \cdot 3) [\;m^2 + (2 \cdot 3)\;]\,[\mu 1 2 Q_0]_\lambda - 
     2\,(2 \cdot 3) [\mu 1 2 Q_0 3 1]_\lambda  
   \right.
   \nonumber \\ &&
   +  2\,(1 \cdot 2) \{\;2\,(2 \cdot 3) [\mu 1 2 Q_0]_\lambda 
- 2\,(2 \cdot 3) [\mu 1 Q_0 2]_\lambda + 
        [\mu 1 2 Q_0 3 1]_\lambda + [\mu 2 1 3 2 Q_0]_\lambda\; \}  
    \nonumber \\ && \left.
   + [\mu 1 Q_0 2 3 1 2 Q_0]_\lambda - [\mu Q_0 1 3 2 1 Q_0 2]_\lambda + 
     [\mu Q_0 1 3 2 Q_0 1 3]_\lambda \right\}
  \nonumber \\ &&
 -8\,(D_0 - D_2) (1 \cdot 2) (2 \cdot 3) 
   \left\{2\,(1 \cdot 3) \{[\;m^2 + (2 \cdot 3)\;]\,[\mu 1 3
   Q_0]_\lambda 
- (1 \cdot 2) [\mu 2 3 Q_0]_\lambda\}
   \right.
   \nonumber \\ &&
    \left.
    + 
     (1 \cdot 2) [\;4\,(2 \cdot 3) [\mu 1 3 Q_0]_\lambda 
+ [\mu 1 3 Q_0 3 1]_\lambda\;] - 
     (2 \cdot 3) [\;[\mu 1 3 Q_0 2 1]_\lambda + [\mu 1 3 Q_0 3 1]_\lambda\;]
     \right\}
\nonumber \\ &&
 -2\,(D_0 - D_{-1}) (2 \cdot 3) \left\{4\,(1 \cdot 2)^2 
[\mu 3 Q_0 2 3 1]_\lambda + 
     m^2 [\mu 2 1 3 Q_0 2 3 1]_\lambda - m^2 [\mu 2 3 1 2 Q_0 1 3]_\lambda
   \right.
   \nonumber \\ &&  
      + 
     2\,(2 \cdot 3) [\mu Q_0 1 3 Q_0 2 1 3]_\lambda - 
     2\,(1 \cdot 2) \{4\,(2 \cdot 3)^2 [\mu 1 2 Q_0]_\lambda 
      +  4\,m^2 (1 \cdot 3) [\mu 3 2 Q_0]_\lambda  
 \nonumber \\ &&     
       + 2\,(2 \cdot 3) [\;2\,(1 \cdot 3) [\mu 3 2 Q_0]_\lambda 
- [\mu 2 3 1 2 Q_0]_\lambda - 
           [\mu 3 2 1 Q_0 2]_\lambda + [\mu 3 Q_0 2 3 1]_\lambda - 
           [\mu Q_0 1 3 Q_0 2]_\lambda\;] 
  \nonumber \\ &&         
           + [\mu 3 Q_0 2 3 1 2 Q_0]_\lambda 
        + [\mu Q_0 3 1 Q_0 2 3 1]_\lambda \}   
         + [\mu 2 1 3 2 1 Q_0 2 3 1]_\lambda + 
     [\mu 2 3 1 Q_0 2 3 1 2 Q_0]_\lambda 
\nonumber \\ && \left.
     -[\mu Q_0 1 3 Q_0 2 3 1 2 Q_0]_\lambda 
     - [\mu Q_0 3 1 Q_0 2 1 3 2 Q_0]_\lambda
\right\}
\nonumber \\ &&
   -8\,(D_0 + m^2) (1 \cdot 2) (2 \cdot 3) 
   \left\{
   4\,(1 \cdot 2) (2 \cdot 3) \{(1 \cdot 3) ([\mu 1]_\lambda 
    - [\mu 2]_\lambda)
    + [\mu 1 3 2]_\lambda\}  
    \right.
    \nonumber \\ && \left.
        +  (1 \cdot 3) \left[\;
           -4\,(2 \cdot 3)^2 [\mu 1]_\lambda 
           + 2\,(2 \cdot 3) [\mu 1 3 2]_\lambda + 
           m^2 (2\,[\mu 1 3 2]_\lambda - [\mu 2 3 1]_\lambda)
                       \;\right]
                    \right\}
\nonumber \\ && 
  + 4\,(2 \cdot 3) \left\{m^2 (1 \cdot 3) [\mu 2 3 1 Q_0 2 1 3]_\lambda + 
     2\,(1 \cdot 2)^2 \{
   4\,(2 \cdot 3)^2 ([\mu 1 2 Q_0]_\lambda - [\mu 1 Q_0 2]_\lambda) 
   \right.
   \nonumber \\ &&   
      +  2\,m^2 (1 \cdot 3) ([\mu 2 3 1]_\lambda + [\mu 3 2 Q_0]_\lambda) 
       + 2\,(2 \cdot 3) ([\mu 1 2 Q_0 3 1]_\lambda 
       + [\mu 2 1 3 2 Q_0]_\lambda) + [\mu 2 3 1 Q_0 2 3 1]_\lambda
                      \} 
   \nonumber \\ &&    
   + (1 \cdot 2) \{
   -4\,(2 \cdot 3)^2 [\;[\mu 1 2 Q_0 3 1]_\lambda
                      -2\,(1 \cdot 3)[\mu 1 2 Q_0]_\lambda \;]  
  \nonumber \\ &&  
       + 2\,(1 \cdot 3) \left[
           2\,m^2 (2 \cdot 3) (2\,[\mu 1 2 Q_0]_\lambda 
              + [\mu Q_0 1 2]_\lambda + 
              [\mu Q_0 1 3]_\lambda) \right. 
   \nonumber \\ &&   \left. \left.   
          + m^2 (m^2 [\mu 2 3 1]_\lambda - [\mu 1 Q_0 2 3 1]_\lambda + 
              [\mu 2 3 1 Q_0 2]_\lambda)
                         \right] + 
        m^2 ([\mu 2 3 1 2 Q_0 1 3]_\lambda + [\mu 2 3 1 Q_0 2 3
   1]_\lambda)
                 \}
        \right\}\,. \nonumber \\
&&
\eea
The expressions for $I^\mu_{2}$ and $I^\mu_{3}$ 
are given in terms of $I^\mu_{1}$ in \mbox{eqs. (\ref{i2exp}) 
and (\ref{i3exp})}.

Another advantage of the above reduction technique is the possibility
of checking the calculation without explicitly computing the loop integrals.
For example, with any arbitrary choice of $p_1, p_2, p_3, q$ and $m_e
\equiv m$, the expressions in \mbox{eqs. (\ref{exam}) and (\ref{apmm2})}
should coincide numerically.

\clearpage
\begin{table}
\begin{tabular}{|c|c|c|c|}
\hline
$\omega/m_e$&$\gamma\nu\to\gamma\gamma\nu$&$\gamma\gamma\to\gamma\nu\bar\nu$&$\nu\bar\nu\to\gamma\gamma\gamma$ \\ \hline 
0.3 &0.969(8)  &1.09(1)  &1.20(1)  \\ \hline
0.4 &0.923(6)  &1.17(1)  &1.37(1)  \\ \hline
0.5 &0.888(6)  & 1.28(1) &1.68(1)  \\ \hline
0.6 &0.852(4)  &1.47(1)  &2.20(1)  \\ \hline
0.7 &0.826(5)  &1.80(1)  &3.17(1)  \\ \hline
0.8 &0.811(5)  &2.41(1)  &5.31(2)  \\ \hline
0.9 &0.819(6)  & 3.95(2) &11.88(3) \\ \hline
1.0 & 0.880(7) &23.1(1)  &176.3(3) \\ \hline
1.1 &1.19(1)   &18.2(1)  &94.7(2)  \\ \hline
1.3 &1.71(2)   &8.31(5)  &31.6(1)  \\ \hline
1.5 &1.44(1)   &3.37(2)  &11.9(1)  \\ \hline
1.7 &0.996(8)  &1.40(1)  &4.96(3)  \\ \hline
1.9 &0.635(4)  &0.622(3) &2.23(1)  \\ \hline
2.0 &0.503(3)  &0.424(2) & 1.54(1) \\ \hline
\end{tabular}
\vspace{1cm}
\caption{{Ratio between exact and effective results for 
the three cross sections. 
The error on the last digit comes from the phase-space Monte Carlo 
integration.}}\label{tab-exact}
\end{table}

\setcounter{figure}{1}
\begin{figure}[h]   

\hspace{1.5cm}\huge{$\sigma(\gamma\nu\to\gamma\gamma\nu)$ (fb)}\\
\vspace{-5cm}

$
\epsfbox{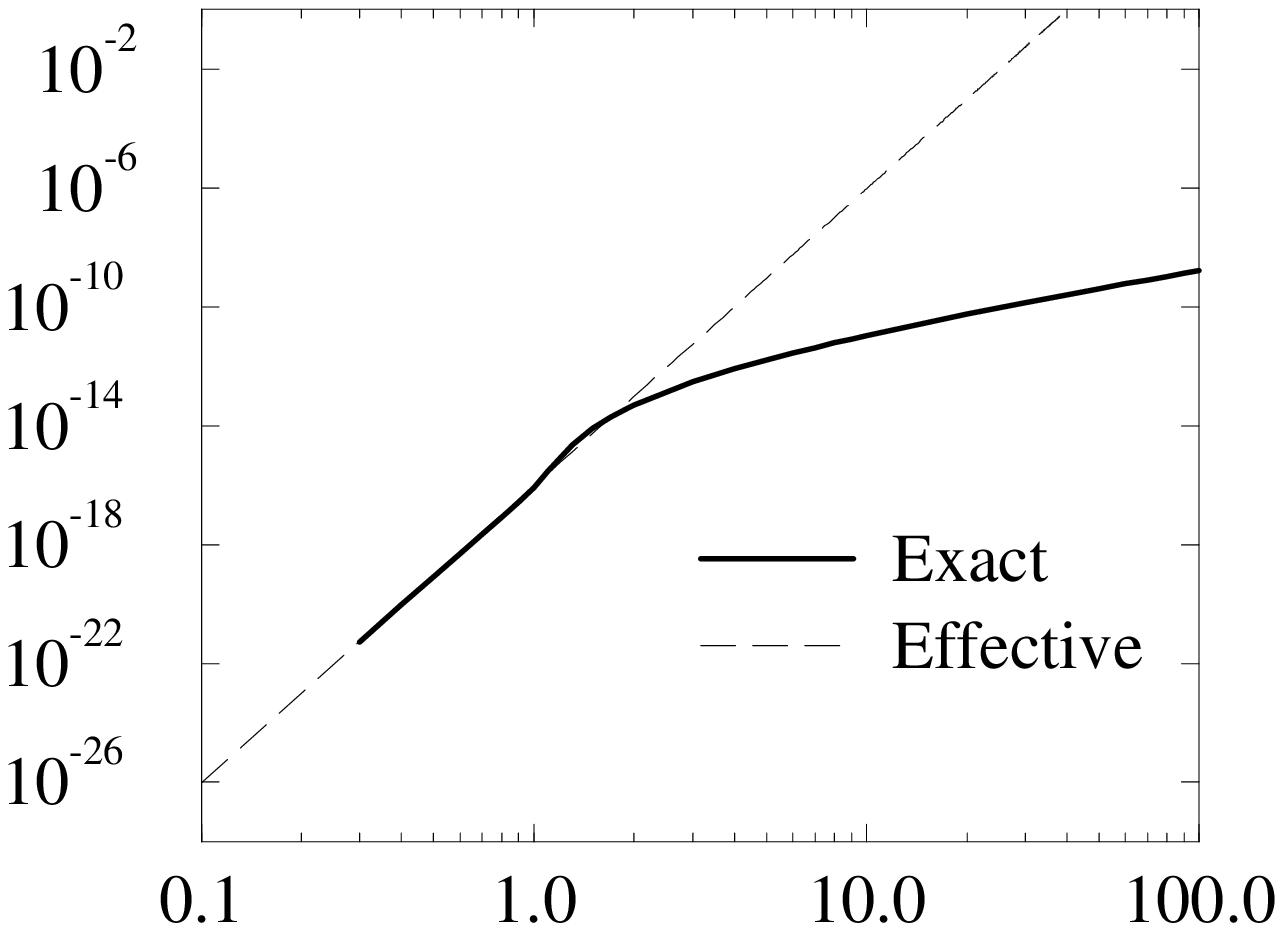}
$

\vspace{-2cm}
\hspace{11cm} 
\huge{$\omega/m_e$} 
\vspace{2cm}
\caption[]{\large{$\gamma\nu\to\gamma\gamma\nu$ cross section in fb 
as a function of $\omega/m_e$.}} \protect\label{complet1}
\end{figure}

\newpage
\begin{figure}[h]   

\hspace{1.5cm}\huge{$\sigma(\gamma\gamma\to\gamma\nu\bar\nu)$ (fb)}\\
\vspace{-5cm}

$
\epsfbox{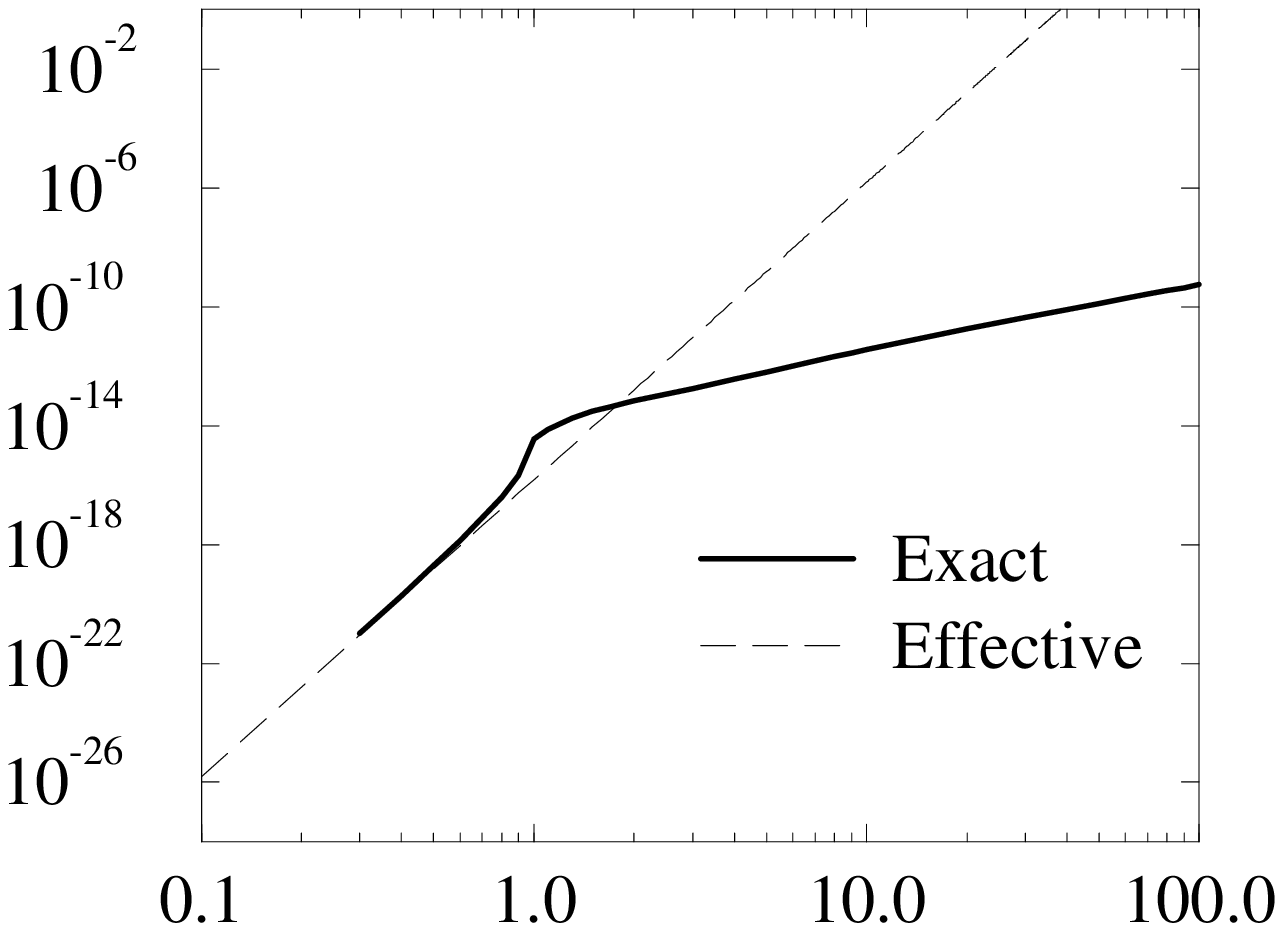}
$

\vspace{-2cm}
\hspace{11cm} 
\huge{$\omega/m_e$} 
\vspace{2cm}
\caption[]{\large{$\gamma\gamma\to\gamma\nu\bar\nu$ cross section in fb 
as a function of $\omega/m_e$.}} \protect\label{complet2}
\end{figure}

\newpage

\begin{figure}[h]   

\hspace{1.5cm}\huge{$\sigma(\nu\bar\nu\to\gamma\gamma\gamma)$ (fb)}\\
\vspace{-5cm}

$
\epsfbox{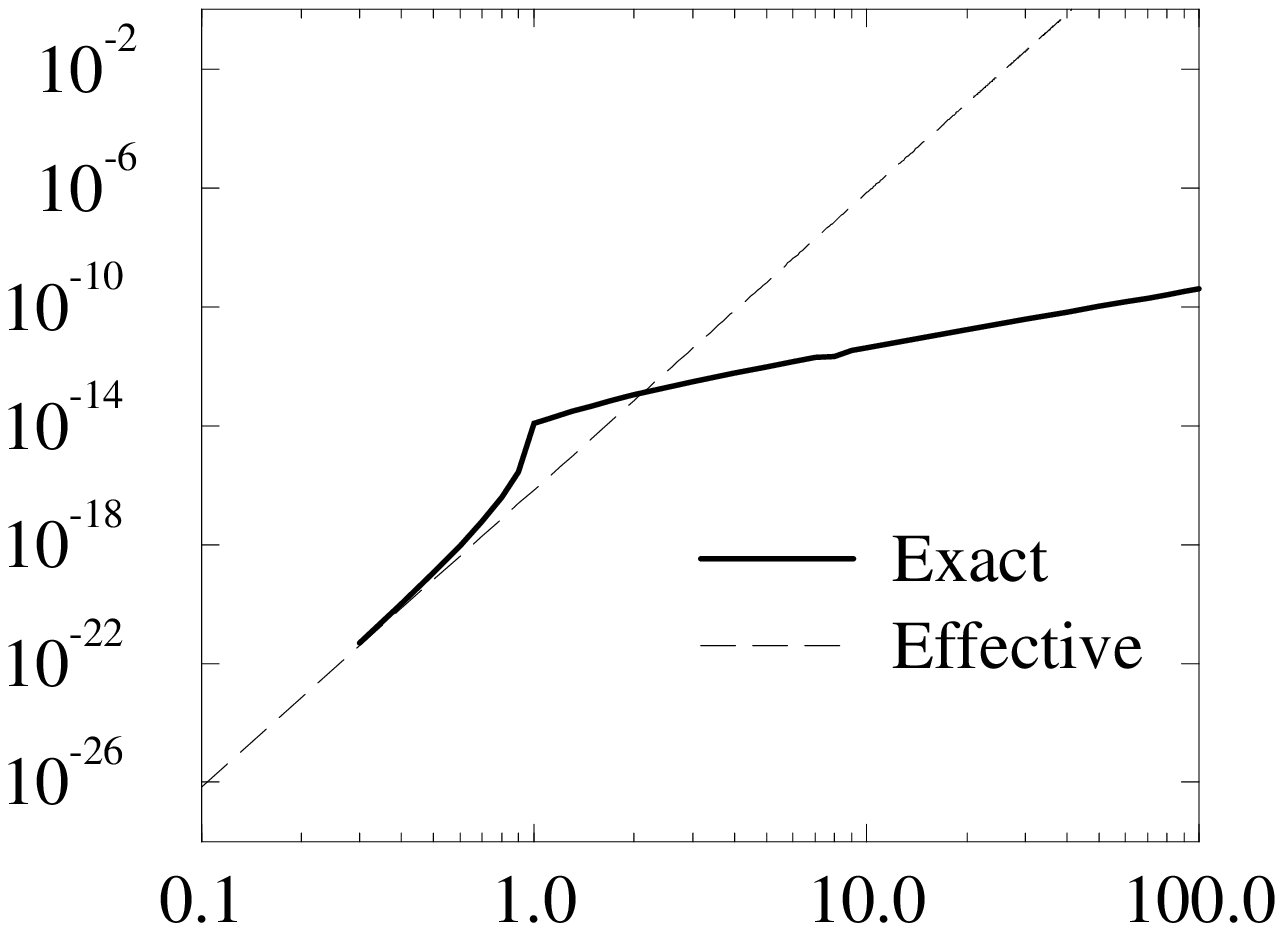}
$

\vspace{-2cm}
\hspace{11cm} 
\huge{$\omega/m_e$} 
\vspace{2cm}
\caption[]{\large{$\nu\bar\nu\to\gamma\gamma\gamma$ cross section in
    fb as a function of $\omega/m_e$.}} \protect\label{complet3}
\end{figure}
\end{document}